 quant-ph/0401025

 \documentclass[prl,aps,showpacs]{revtex4}
 \usepackage{graphicx}

\begin{document}

 \def\BE{\begin{equation}}
 \def\EE{\end{equation}}
 \def\BEA{\begin{eqnarray}}
 \def\EEA{\end{eqnarray}}
 \def\BA{\begin{array}}
 \def\EA{\end{array}}
 \def\O{\Omega}

 \title{Squeezed-light source for the superresolving microscopy}
 \author{Ivan V.~Sokolov$^{1}$ and Mikhail I.~Kolobov$^{2}$}
 \affiliation{1 V.A.Fock Physics Institute, St.-Petersburg University,
       198504 Stary Petershof, St.-Petersburg, Russia}
 \affiliation{2 Laboratoire PhLAM, Universit\'e de Lille-1,
       F-59655 Villeneuve d'Ascq cedex, France}

\begin{abstract}
We propose a source of multimode squeezed light that can be used
for the superresolving microscopy beyond the standard quantum
limit. This source is an optical parametric amplifier with a
properly chosen diaphragm on its output and a Fourier lens. We
demonstrate that such an arrangement produces squeezed prolate
spheroidal waves which are the eigen modes of the optical imaging
scheme used in microscopy. The degree of squeezing and the number
of spatial modes in illuminating light, necessary for the
effective object field reconstruction, are evaluated.
\end{abstract}

\pacs{42.50.Dv,42.30.Wb, 42.50.Lc}

\maketitle

Spatial behavior of nonclassical light is presently attracting an
increasing interest both in theory and experiment
\cite{Kolobov99,EPJD}. Quantum effects in optical imaging and
other transverse spatial phenomena are studied within a framework
of a European project "Quantum Imaging" \cite{QUANTIM}. One of the
problems recently addressed in this context is about the ultimate
quantum limit of the optical resolution \cite{Kolobov00}. A
classical criterion of resolution was formulated at the end of the
last century  by Abbe and Rayleigh. It states that optical
resolution is limited by diffraction on the system pupil. However,
though it might be difficult to accept, the diffraction limit is
not a fundamental limit like, for example the Heisenberg
uncertainty relation. Recent publications have theoretically
demonstrated that in principle this limit can be beaten for both
writing and reading of optical information.

Sub-diffraction-limited optical recording of information, coined
as "quantum lithography", uses a specially designed nonclassical
light, $N$-photon entangled state, together with a $N$-photon
absorber which allows to reach theoretically a resolution of
$\lambda/N$ \cite{Boto00,Kok02}. To go beyond the diffraction
limit in read-out of optical information like in imaging, is in
principle possible having some a priori information about the
object using the so-called "superresolution" techniques. As shown
recently in Ref.~\cite{Kolobov00} the ultimate limit in such
superresolution is set not by diffraction but by the quantum
fluctuations of light illuminating the object of finite size and
by the vacuum fluctuations outside the object. These quantum
fluctuations of the illuminating light together with the vacuum
fluctuations outside the object set up {\it the standard quantum
limit of superresolution} which can be much smaller than the
diffraction limit. Moreover, one can go beyond this standard
quantum limit using a specially designed nonclassical light for
illumination of the object. In recent experiment \cite{Treps02} it
was demonstrated that using multimode squeezed light allows to
measure a displacement of 2.9 {\AA} of a laser beam with
wavelength $\lambda=1064$ nm.

In Ref.~\cite{Kolobov00} Kolobov and Fabre  have suggested a
scheme that allows to improve the superresolution beyond the
standard quantum limit in reconstruction of an optical object
using a multimode squeezed light. This scheme was formulated in
terms of the so-called prolate spheroidal waves which are the
eigenfunctions of the optical imaging scheme. To achieve the
superresolution with multimode squeezed light one has to prepare
these prolate waves in squeezed state. The question remained on
how to produce such squeezed prolate waves.

In this letter we provide the answer to this question. Precisely, we
demonstrate that an optical parametric amplifier (OPA) with a properly
chosen diaphragm on its output and a Fourier lens, produces squeezed
prolate spheroidal waves used in the superresolving microscopy. We
investigate the quantum statistics of the squeezed prolate spheroidal
waves in our scheme in dependence of the physical parameters of the OPA
and of the optical configuration. We formulate simple estimates on the
number of the "object elements" to be reconstructed in connection with the
number of degrees of freedom in the nonclassical illuminating light.

The scheme of optical imaging with multimode squeezed light is
shown in Fig.~1. For simplicity we consider a one-dimensional
case.
\begin{figure}[h]
\includegraphics[width=100mm]{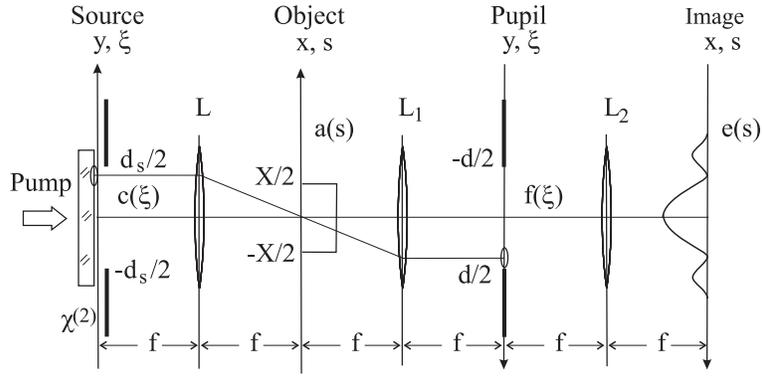}
\caption{Schematic of the optical imaging with squeezed light.}
\end{figure}
The part of the scheme to the right of the object plane performs
diffraction-limited imaging of an object of finite size $X$
located in the object plane \cite{Kolobov00}. The first lens $L_1$
performs the Fourier transform of the object field into the pupil
plane with a pupil of size $d$. After the second Fourier transform
by the lens $L_2$ an image is created in the image plane. This
image is a diffraction-limited copy of the object due to the
finite size of the pupil.

The part to the left of the object plane is an illumination
scheme. It consists of a traveling-wave OPA placed in the source
plane and a Fourier lens $L$. It is well-known from the literature
that a traveling-wave OPA with plane-wave pump and nonlinear
crystal with large transverse area creates multimode squeezed
vacuum on its output \cite{Kolobov99}. A new feature of our scheme
is a diaphragm of size $d_s$ on the output of the OPA which serves
for selection of the transverse modes in squeezed state. As we
demonstrate below, when the size of this diaphragm matches the
size of the pupil, $d_s \geq d$, this setup squeezes exactly the
prolate spheroidal waves which are the eigen modes of the imaging
scheme. This result can be easily understood qualitatively.
Indeed, when all three lenses in the scheme have the same focal
distance $f$, as in Fig.~1, the lenses $L$ and $L_1$ create a
geometrical image of the diaphragm $d_s$ in the pupil plane.
Therefore, it is intuitively clear that one has to match the
diaphragm size $d_s$ and the pupil size $d$ to select the modes of
the source which will pass through the imaging scheme.

Let us introduce the dimensionless coordinates in the object and
the image plane as $s = 2x/X$, and the dimensionless coordinates
in the source and the pupil plane as $\xi=2y/d$ (see Fig.~1). The
dimensionless photon annihilation operators in the source, object,
pupil, and image planes are denoted respectively as ${\hat
c}(\xi), {\hat a}(s), {\hat f}(\xi)$, and ${\hat e}(s)$. These
operators obey the standard commutation relations,
 \BE
      [{\hat c}(\xi),{\hat c}^\dag(\xi')] = \delta(\xi-\xi'),\quad
      [{\hat a}(s),{\hat a}^\dag(s')] = \delta(s-s'),
 \label{free_field}
 \EE
and similar for ${\hat f}(\xi)$, and ${\hat e}(s)$. Naturally, the
annihilation and creation operators at different planes commute
with each other. These operators are normalized so that $\langle
{\hat c^{\dag}}(\xi){\hat c}(\xi)\rangle$, for example, gives the
mean photon number per unit dimensionless length in the source
plane. Similar normalization is valid for the other planes. The
Fourier transform $\left(T \hat{c} \right)(s)$ physically
performed by the lens $L$ in Fig.~1 reads as follows,
 \BE
    {\hat a}(s) = \left(T \hat{c} \right)(s) =
    \sqrt{\frac{c}{2\pi}}\int_{-\infty}^\infty d\xi e^{-i cs\xi}
    {\hat c}(\xi),
    \label{Fourier}
 \EE
where $\displaystyle{c=\frac{\pi}{2}\frac{dX}{\lambda f}}$ is the
space-bandwidth product of the imaging system.

The quantum theory of the imaging scheme in Fig.~1 was formulated
in Ref.~\cite{Kolobov00} in terms of the prolate spheroidal
functions $\psi_k(s)$ \cite{Slepian61,Frieden71}(for some examples
see Ref.~\cite{Kolobov01}). These are the eigen functions of the
imaging operator of the scheme, orthonormal on the interval
$-\infty <s< \infty$. The photon annihilation operator ${\hat
a}(s)$ in the object plane can be written as decomposition over
$\psi_k(s)$,
 \BE
      {\hat a}(s) = \sum_{k=0}^{\infty} {\hat A}_k \psi_k(s) + {\hat N}(s),
       \label{expansion}
 \EE
where the operator-valued coefficients ${\hat A}_k$ are evaluated
as
 \BE
      {\hat A}_k = \int_{-\infty}^\infty ds\,{\hat a}(s)\psi_k(s).
       \label{expansion_coefficients_def}
 \EE
The operators ${\hat A}_k$ and ${\hat A}^{\dag}_k$ satisfy the
standard commutation relations of the photon annihilation and
creation operators for discrete modes,
 \BE
      [{\hat A}_k,{\hat A}_{k'}^\dag] = \delta_{k,k'}, \quad
      [{\hat A}_k,{\hat A}_{k'}] = 0.
        \label{commutation}
 \EE
The Fourier transform ${\tilde\psi}_k(\xi)$ of the prolate
spheroidal functions $\psi_k(s)$, performed by the lens $L_1$, is
zero outside the interval $|\xi|\leq 1$. Consequently, the set of
functions $\{\psi_k(s)\}$ is not complete in the Hilbert space
$L^2(-\infty,\infty)$, and to satisfy the commutation relations
(\ref{free_field}) one has to add an additional term ${\hat
N}(s)$. This term has zero Fourier spectrum in the interval
$|\xi|\leq 1$ and does not contribute to the coefficients ${\hat
A}_k$,
 \BE
      \int_{-\infty}^\infty ds \,\psi_k(s) {\hat N}(s) = 0.
      \label{orthogonal}
 \EE
The field $\hat{N}(s)$ can be decomposed over the complementary set of
prolate functions $\{\theta_k(s)\}$, orthogonal to $\{\psi_k(s)\}$.

Physically speaking, the first term in Eq.~(\ref{expansion}) is
the object field component that propagates in our scheme through
the pupil to the image plane. The second object field component
$\hat{N}(s)$ is absorbed outside the pupil and is not observed.
Therefore, in what follows we shall omit ${\hat N}(s)$ in the
object field (\ref{expansion}).

To obtain the canonical transformation of the photon annihilation
and creation operators in the imaging scheme on Fig.~1, one has to
split the coordinates $s$ and $\xi$ into two regions, the "core",
$|s|\leq 1$ and $|\xi|\leq 1$, corresponding to the central
regions of the object (image) and the source (pupil), and the
"wings", $|s|> 1$ and $|\xi|> 1$, outside these areas. The
orthonormal bases in these regions of the object (image) plane are
given by
 \BE
    \varphi_k(s) = \left\{\begin{array}{cl}
    {\displaystyle\frac{1}{\sqrt{\lambda_k}}\psi_k(s)} & \quad  |s|\leq 1,\\
    0 & \quad  |s|>1,\\
    \end{array} \right. \quad\quad\quad
    \chi_k(s) = \left\{\begin{array}{cl}
    0 & \quad |s|\leq 1,\\
    {\displaystyle\frac{1}{\sqrt{1-\lambda_k}}\psi_k(s)} & \quad  |s|>1,\\
    \end{array} \right.
           \label{varphi_chi}
 \EE
where $\lambda_k$ are the eigenvalues of the corresponding prolate
spheroidal functions $\psi_k(s)$, depending on the space-bandwidth
product $c$. We should note that the functions $\varphi_k(s)$ are
complete in the Hilbert space $L^2(-1,1)$.

It follows from (\ref{varphi_chi}) that the functions $\psi_k(s)$
and $\theta_k(s)$ can be written in the form
 \BE
 \psi_k(s) = \sqrt{\lambda_k} \varphi_k(s) + \sqrt{1-\lambda_k}  \chi_k(s),
 \quad
 \theta_k(s) = \sqrt{1-\lambda_k} \varphi_k(s) - \sqrt{\lambda_k}
 \chi_k(s).
 \label{psi_theta}
 \EE
Similar relations take place in the source (pupil) plane.

In terms of two sets $\{\varphi_k(s)\}$ and $\{\chi_k(s)\}$ we can
write the annihilation operators in the source and the object
planes as
 \BE
     {\hat c}(\xi) = \sum_{k=0}^{\infty} {\hat c}_k \varphi_k(\xi) +
     \sum_{k=0}^{\infty} {\hat d}_k \chi_k(\xi),
              \label{source_plane}
 \EE
and
 \BE
     {\hat a}(s) = \sum_{k=0}^{\infty} {\hat a}_k \varphi_k(s) +
     \sum_{k=0}^{\infty} {\hat b}_k \chi_k(s).
              \label{object_plane}
 \EE
Here ${\hat c}_k$ and ${\hat a}_k$ are the annihilation operators
of the prolate modes $\varphi_k$ in the core region of the source
and the object planes, while ${\hat d}_k$ and ${\hat b}_k$ are the
annihilation operators of the prolate modes $\chi_k$ in the wings
regions. The operators ${\hat c}_k$ and ${\hat d}_k$ are expressed
through the field operator ${\hat c}(\xi)$ by
 \BE
      {\hat c}_k = \int_{-\infty}^\infty d\xi\,{\hat c}(\xi)\varphi_k(\xi),
      \quad\quad
      {\hat d}_k = \int_{-\infty}^\infty d\xi\,{\hat c}(\xi)\chi_k(\xi).
       \label{coefficients_core_wings}
 \EE
Similar relations hold for ${\hat a}_k$, ${\hat b}_k$ and ${\hat a}(s)$.

In our analysis we shall use the following property of prolate
spheroidal functions \cite{Frieden71},
 \BE
 \int_{-1}^{1}d\xi \psi_k(\xi) e^{-iq\xi} =
 (-i)^k\sqrt{\frac{2\pi\lambda_k}{c}} \psi_k(q/c).
 \label{prolate_property}
 \EE
Using (\ref{varphi_chi}), (\ref{psi_theta}), the field transform
(\ref{Fourier}) between the source and the object plane and this equation,
one can find the following propagation relations for the core and wings of
the light wave emitted by the source:
 $$
 \left(T\varphi_k\right)(s) = (-i)^k\left[\sqrt{\lambda_k}
 \varphi_k(s) + \sqrt{1-\lambda_k} \chi_k(s)\right] = (-i)^k
 \psi_k(s),
 $$
 \BE
 \left(T\chi_k\right)(s) = (-i)^k\left[\sqrt{1-\lambda_k}
 \varphi_k(s) - \sqrt{\lambda_k} \chi_k(s)\right] = (-i)^k
 \theta_k(s).
 \label{propagation}
 \EE
Taking into account (\ref{source_plane}) and (\ref{object_plane})
we obtain,
 \BE
 \hat{a}(s) =\left(T\hat{c}\right)(s) =
 \sum_{k=0}^{\infty} (-i)^k \left[ {\hat c}_k \psi_k(s) +
 {\hat d}_k \theta_k(s) \right],
 \label{propagation_a}
 \EE
and
 \BE
  {\hat A}_k = (-i)^k{\hat c_k}.
  \label{A_k}
 \EE
As expected, there is no contribution into ${\hat A}_k$ from the
second sum in Eq.~(\ref{source_plane}) containing operators ${\hat
d}_k$ and describing the illumination coming from the wings area
of the source. If the source diaphragm is larger or matches the
size of the pupil, $d_s \geq d$, it has no effect on the operator
amplitudes ${\hat c}_k$ and ${\hat A}_k$, see
(\ref{coefficients_core_wings}), and we shall neglect the
diaphragm in the calculation of $\hat{A}_k$.

To obtain explicitly the coefficients ${\hat A}_k$ in case of
illumination of the scheme by a traveling-wave OPA we shall use
simplified description of an OPA with a plane-wave undepleted pump
(see Ref.~\cite{Kolobov99}). In this approximation one can find
analytically the spatial Fourier amplitudes ${\tilde c(q)}$,
 \BE
    {\tilde c}(q) = \int_{-\infty}^\infty d\xi e^{-i q\xi}
    {\hat c}(\xi),
    \label{Fourier_coeff}
 \EE
of the field at the output of the crystal as a linear
transformation of the corresponding input Fourier amplitudes
${\tilde c}_{in}(q)$ and ${\tilde c}^{\dag}_{in}(q)$,
 \BE
    {\tilde c}(q) = U(q) {\tilde c}_{in}(q) + V(q) {\tilde c}_{in}^\dag(-q).
    \label{squeezing}
 \EE
Here the operators ${\tilde c}_{in}(q)$ at the input of the OPA
are in the vacuum state. The complex coefficients $U(q)$ and
$V(q)$ depend on the nonlinear susceptibility of the crystal, its
length, and the matching conditions in the OPA. These coefficients
have the property,
 \BE
      |U(q)|^2 - |V(q)|^2 = 1,
      \label{unitarity}
 \EE
that guarantees the preservation of the commutation relations (\ref{free_field}).
Using the standard parameters of multimode squeezing (see Ref.~\cite{Kolobov99})
we obtain:
 \BE
      U(q) \pm V^*(-q) = e^{-i\varphi(q)} \left[ e^{\pm r(q)} \cos\theta(q)
      + ie^{\mp r(q)} \sin\theta(q) \right].
            \label{U_and_V}
 \EE
The phase of the amplified (stretched) quadrature amplitude of the
OPA output field ${\tilde c}(q)$ is $\theta(q)$. The phase factor
$\exp(-i\varphi(q))$ specifies the phase of the quadrature
amplitudes of the input field ${\tilde c}_{in}(q)$ which are
squeezed or stretched. For the vacuum input field of the OPA the
last phase is irrelevant.

The operator amplitudes ${\hat A}_k$ are found explicitly with the
use of Eqs.~(\ref{A_k}) and (\ref{coefficients_core_wings}). The
quantities ${\hat c}_{in}(\xi)$ and $\varphi_k(\xi)$ are expressed
through their Fourier transforms (\ref{squeezing}) and
(\ref{propagation}). After some calculation we obtain,
 \BE
      {\hat A}_k = \frac{1}{\sqrt{2\pi c}}
      \int_{-\infty}^\infty dq\, \psi_k(q/c) \left[U(q) {\tilde c}_{in}(q) +
      V(q) {\tilde c}_{in}^\dag(-q)\right].
          \label{expansion_coefficients_final}
 \EE
We shall introduce the real quadrature components of the field
amplitudes ${\hat A}_k$ in the object plane as
 \BE
     {\hat A}_k={\hat A}_{1k}+i{\hat A}_{2k}.
          \label{quadratures}
 \EE
For the variances of these quadrature components we obtain
 \BE
      \langle \left(\Delta {\hat A}_{1k}\right)^2 \rangle = \frac{1}{4c}
      \int_{-\infty}^\infty dq\, \psi_k^2(q/c)\left[ e^{\pm 2r(q)}
      \cos^2\theta(q) +  e^{\mp 2r(q)} \sin^2\theta(q)\right],
         \label{prolate_variance_1}
 \EE
 \BE
      \langle \left(\Delta {\hat A}_{2k}\right)^2 \rangle = \frac{1}{4c}
      \int_{-\infty}^\infty dq\, \psi_k^2(q/c)\left[ e^{\mp 2r(q)}
      \cos^2\theta(q) +  e^{\pm 2r(q)} \sin^2\theta(q)\right].
         \label{prolate_variance_2}
 \EE
Here the upper and the lower sign correspond respectively to the even
and the odd prolate spheroidal functions $\psi_k$.

It follows from this result that the prolate spheroidal waves in the
object illumination can be prepared in squeezed state. By
proper choice of the squeezing phase $\theta(q)$ at low spatial
frequencies $q$ one can minimize quantum fluctuations in one of the
quadrature amplitudes $A_{\sigma k}$, $\sigma = 1, 2$ (namely, in the one
detected in the image plane of our scheme). Taking the degree and
the phase of squeezing as constant, $r(q) = r$, $\theta(q) =
\theta$, we can estimate the variance of the squeezed
quadrature amplitude:
 \BE
      \langle \left(\Delta A_{\sigma k}\right)^2 \rangle \sim e^{-2r}/4.
        \label{estimate_variance}
 \EE
In reality the spatial-frequency band of multimode squeezing is
limited by the phase-matching condition in the OPA. One has to
take into account the spatial-frequency dispersion of squeezing,
that is, the frequency dependence of the squeezing phase $\theta(q)$
due to diffraction in free space and inside the OPA. Both
phenomena deteriorate squeezing of prolate spheroidal waves.

As shown in \cite{Kolobov89,Kolobov99,Sokolov99}, the effect of
diffraction on squeezing can be almost perfectly compensated by means of
adjustment of the lens array. If this is done, it follows from
(\ref{prolate_variance_1}), (\ref{prolate_variance_2}), that the degree of
squeezing of prolate spheroidal waves depends on the overlap in the object
plane of two areas: (i) the area illuminated by the squeezed plane waves
${\tilde c}(q)$, which are focused by lens $L$ to the points $s = q/c$,
and (ii) the area of the energy localization $\sim \psi_k^2(s)$ of the
prolate spheroidal waves.

In analogy to other phenomena with spatially-multimode squeezed
light, in the even and odd components of the object field the
different quadrature amplitudes are squeezed; for, say, $\theta =
0$, the squeezed quadratures are $A_{1k}$ and $A_{2k}$ for the odd
and the even prolate waves respectively. In measurement this
implies, e.g., the spatially resolved homodyne detection with
numerical evaluation of the odd or even squeezed amplitudes
(depending on the local oscillator phase).

Finally, we can formulate the conditions on multimode squeezing in
the object illumination in terms of the number $N$ of independent
degrees of freedom in the light field, propagating through the
diaphragm  of the OPA (we assume here the optimum size $d_s = d$
of the diaphragm). The properties of the OPA emission can be
characterized by the coherence length $l_{c}$ of the output field
${\hat c}(\xi)$. The spatial-frequency range $q_c$ of effective
squeezing is related to the coherence length by the estimate
 \BE
      |q_c| \leq \frac{\pi d}{2 l_c}.
      \label{freq_range_1}
 \EE
The minimum requirement on the OPA is that the effectively
squeezed waves ${\tilde c}(q_c)$ illuminate, after passing the
lens $L$, the object region $|s| \leq 1$. That is, the waves
 \BE
      |q_c|/c \geq 1,
      \label{freq_range_2}
 \EE
should be squeezed. This gives the following estimate for the
coherence length,
 \BE
     l_{c} \leq \frac{\pi d}{2c},
      \label{coherence_length}
 \EE
and for the number $N$ of independent degrees of freedom in the
illuminating light, emitted from the region $d$:
 \BE
      N \sim d/l_{c} \geq \frac{d X}{\lambda f} = S.
        \label{number}
 \EE
Here $S$ is the Shannon number of our optical scheme.

As seen from (\ref{psi_theta}), the wave profiles $\psi_k(s)$ can be
expanded in terms of two bases, $\{\varphi_k(s)\}$ and $\{\chi_k(s)\}$,
representing the core and the wings of the illuminating field in the
object plane, where $\lambda_k \leq 1$ are the eigenvalues of the imaging
transformation. As known from the theory of prolate spheroidal functions
\cite{Slepian61,Frieden71}, these eigenvalues are close to 1 only for $k
\leq S$, where $S$ is the Shannon number. For higher values of the index
$k$ the field energy in prolate spheroidal waves $\psi_k(s)$ is
concentrated in the wings, that is, outside the object area $|s| \leq 1$.
Hence, the condition $N \sim S$ provides an effective squeezing only of
the prolate spheroidal waves with $k \leq S$. In order to minimize quantum
noise of the higher prolate spheroidal waves (with $k > S$), it is
necessary to use OPA with large number of effectively squeezed spatial
modes of radiation,
 \BE
      N \gg S,
      \label{k_gg_S}
 \EE
and to illuminate by non-classical light a spot in the object
plane with the size much larger than the object itself.

When the condition $N \gg S$ is met, one can illuminate the object
(within the area $|s| \leq 1$) by a bright classical wave with a
properly chosen phase. Since for $\lambda_k \ll 1$ our imaging
scheme is analogous to a highly non-symmetrical beamsplitter
\cite{Kolobov00}, the squeezed illumination concentrated mainly in
the wings area of the object plane allows for the low-noise
measurement of the relevant amplitudes of the image field.
\\

This work was supported by the Network QUANTIM (IST-2000-26019) of the
European Union and by the INTAS under Project No.~2001-2097.

\end{document}